# Study of spectrally resolved thermoluminescence in Tsarev and Chelyabinsk chondrites with a versatile high-sensitive setup

Alexander Vokhmintsev [1], Ahmed Henaish [1,2], Taher Sharshar [3], Osama Hemeda [2] and Ilya Weinstein [1,4,*]

[1] NANOTECH Center, Ural Federal University, Ekaterinburg, 620002, Russia
[2] Physics Department, Faculty of Science, Tanta University, Tanta 31527, Egypt
[3] Physics Department, Faculty of Science, Kafrelshaikh University, Kafr El-Shaikh, Egypt
[4] Institute of Metallurgy of the Ural Branch of the Russian Academy of Sciences, 620016 Ekaterinburg, Russia
* Correspondence: i.a.weinstein@urfu.ru; Tel.: (+7 343 3759374)

**Abstract:** Thermoluminescence (TL) research provides a powerful tool for characterizing radiation-induced processes in extraterrestrial matter. One of the challenges in studying the spectral features of natural TL of stony meteorites is its weak intensity. The present work showcases the capabilities of a high-sensitive original module for measuring the spectrally resolved TL characteristics of the Chelyabinsk and Tsarev chondrites. We have analyzed the emission spectra and glow curves of natural and induced TL over the 300 – 650 nm and RT – 873 K ranges. A quasi-continuous distribution of traps being active within the 350 – 650 K range was found in the silicate substructure of both meteorites under study. Based on the general order kinetic formalism and using the natural TL data, we have also estimated the activation energies of $E_A$ = 0.86 and 1.08 eV for the Chelyabinsk and Tsarev chondrites, respectively.

## 1. Introduction

Thermoluminescence (TL) spectroscopy is a well-proven experimental method for studying the spectral characteristics and kinetic mechanisms of radiation-stimulated processes in irradiated materials [1,2]. In practice, there are different TL-based techniques to apply them in archaeological and geological dating, for dose exposure and radiation contamination monitoring by commercial systems, and to achieve specific targets within solid-state dosimetry using luminescent technologies [3]. Moreover, TL research provides a large-yield stomping ground for characterizing thermal history, describing various impact events, metamorphic processes, and features of the inorganic composition in extraterrestrial materials [4]. Catching of low intensity of natural TL is an arduous enough issue in exploring the spectral properties of meteorites. Usually, an integral luminescent response in a wide wavelength range is only detected. However, this way of studying the meteorites that exhibit simultaneous emission in different bands is not always effective [5,6]. Therefore, one does not succeed in analyzing the spectral peculiarities of the natural and induced TL at a required spectral resolution. Earlier, we have designed and put in practice a high-temperature module as a supplementary unit for commercial fluorescence spectrometers. Besides, its performance was tested within the range of up to 773 K temperature using wide-gap nitrides as examples [7,8]. Along this paper we have demonstrated the capabilities of an original high-temperature module by measuring the spectrally resolved thermoluminescence of the Chelyabinsk and Tsarev meteorites. The goal is to evaluate spectral and energy parameters of thermally stimulated processes in the chondrites exposed by the high-dose irradiation.

## 2. Materials and Methods

Several fragments of the Chelyabinsk LL5 (fall date is 15.02.2013) and Tsarev L5 (fall date is 06.12.1922) chondrites have been studied. The meteorite cores were refined from the fusion crust followed by grinding-up into a micro-sized powder. Further down the line, we held hydrochloric acid treatment of the latter to remove metal particles (see Fig.1).

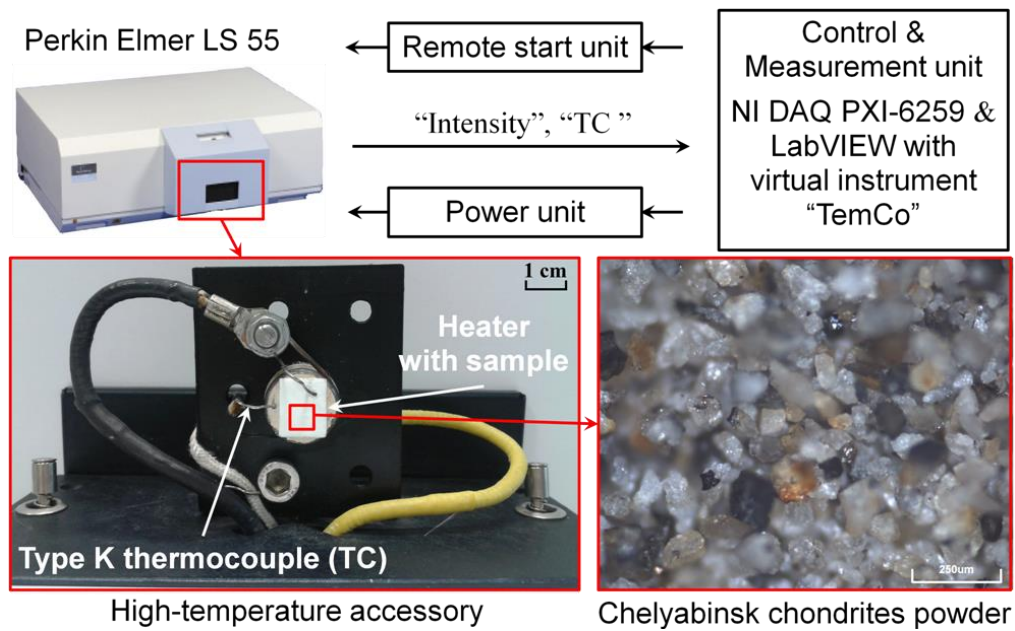

**Figure 1.** Thermoluminescent spectrometer.

The thermoluminescence measurements of the samples were carried out in the phosphorescence regime (12.5 ms gate time and 20 ms cycle time) using a LS 55 Perkin Elmer spectrometer with an original heating accessory module [7]. Figure 1 presents a block diagram of the developed TL spectrometer. The latter includes four main parts such as a high-temperature accessory with a heating stage and a thermocouple, a power unit, a remote start unit, and a control & measurement unit. A detailed configuration and operating regimes of the experimental setup, as well as its abilities to study thermally activated luminescence processes in wide-gap materials were described in Refs. [7,8].

The natural and induced TL glow curves for both meteorites were recorded for the 440 ± 20 nm band within the RT – 873 K range with a linear heating rate of r = 2 K/s. An UELR-10-15S linear accelerator with 10 MeV electrons was utilized for irradiation of the samples and exciting an induced TL response. The radiation doses amounted to 9.1 – 36.4 kGy. For subsequent numerical processing, 4 measurements of TL glow curves were performed for each value of the dose.

The TL spectra ranged at 300 – 650 nm range and RT – 873 K were analyzed with scanning speed of 700 nm/min and r = 0.5 K/s. About 15 spectra have been recorded during single heating process; temperature of the sample changed by 15 K within one measurement of the spectral dependence. In this case, a starting temperature of recording was assigned to each spectrum. Figures with TL spectra and TL glow curves show the selection of the measured dependencies, accounting for the clarity and completeness of presenting the experimental data. The spectral parameters of natural TL for the Tsarev meteorite could not be analyzed due to very low emission intensity.

## 3. Results and Discussion

Figure 2 shows experimental spectra of the natural and induced TL for the Chelyasbinsk and Tsarev chondrites, respectively. A wide structureless band in the visible spectral range is observed in the TL emission for both meteorites under investigation at the indicated temperatures. All the TL spectra can be approximated with high accuracy (coefficient of determination is $R^2 > 0.993$) by a superposition of two $G_1$ and $G_2$ Gaussians, see Figure 3. For the appropriate temperatures, the G1 dominates in the TL emission, its intensity is 4 – 6 times higher than that of the $G_2$ peak. The values of the $E_{max}$ maximum energies and $\omega_E$ half-widths for the Gaussian bands are presented in Table 1 in comparison with independent data on spectral parameters of photo- (PL) and cathodoluminescence (CL) for the same chondrites.

Previous studies of the Chelyabinsk [6] and Tsarev [9] meteorites have shown that the observed PL emission spectra were characterized by two Gaussians also. The shape of the PL bands did not change, while their intensity has decreased with varying excitation photon energy within the 6.2 – 4.5 eV range. The values of spectral parameters obtained for the $G_1$ and $G_2$ components suggest that the PL and TL processes are due to the same recombination centers in both meteorites. In its turn, the CL spectrum for the Chelyabinsk meteorite contains only a single Gaussian-shaped band, its parameters are consistent with the $G_1$ component, see Table 1. It can be concluded that the TL, CL, and PL spectra have the same emission composition, which indicates the similarity of the recombination centers involved into mechanisms of the luminescence of the investigated meteorites exposed by UV, electrons, and space irradiation.

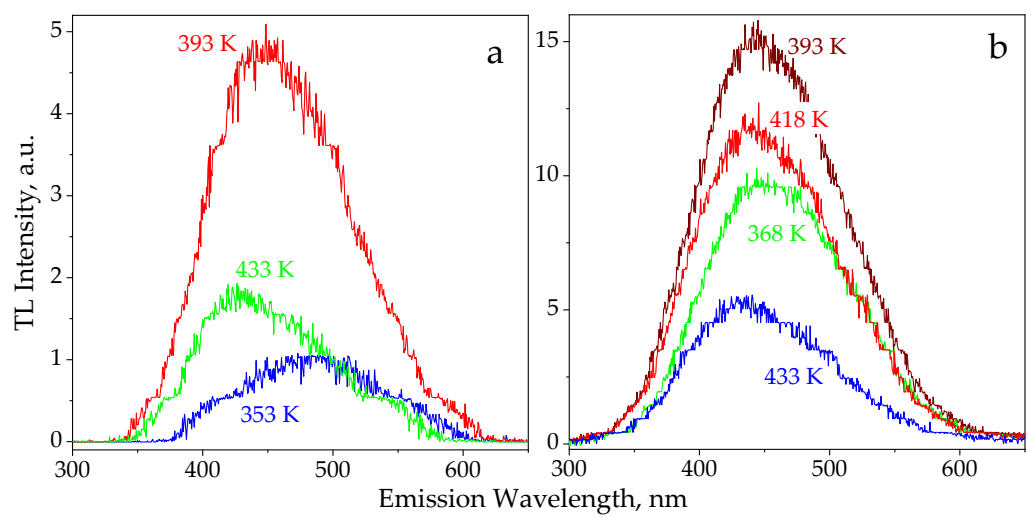

**Figure 2.** Emission spectra of natural TL in the Chelyabinsk (a) and induced TL in the Tsarev (b) chondrites.

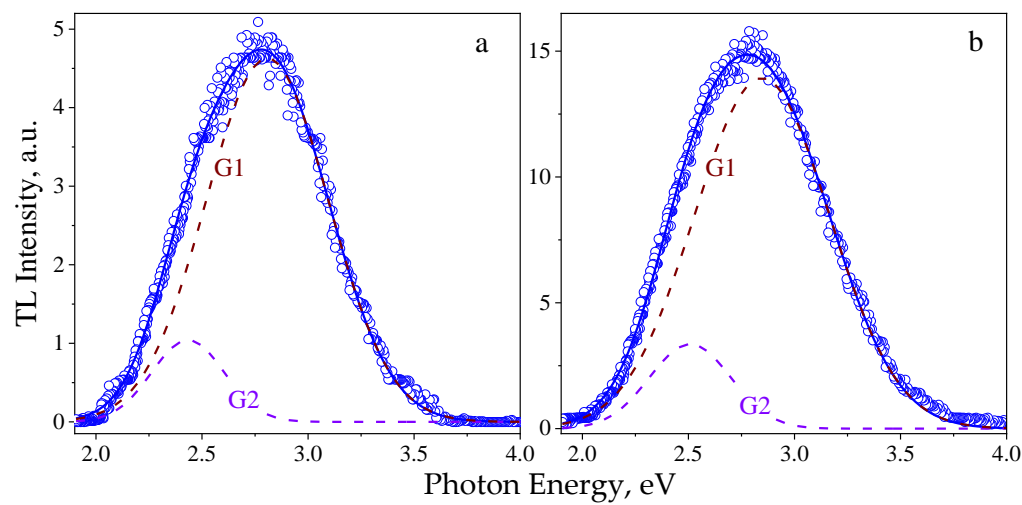

**Figure 3.** Deconvolution of TL spectra measured at 393 K for the Chelyabinsk (a) and Tsarev (b) chondrites. For both plots: circles – experimental data; colored dashed lines – the corresponding Gaussian components; solid blue lines – resulting approximation curves.

Analysis of the data obtained and independent studies allows one to conclude that the observed features of the luminescence can be associated with defective recombination centers in the structure of forsterite [10–12] or enstatite [13]. In the works mentioned above, the shown spectra exhibit a broad band in the 350 – 550 nm range. It should be noted that the 2.75 eV (450 nm) emission is well known for α-quartz and thought to be an intrinsic property of $SiO_4$ tetrahedrons, which are the main structural motifs in the olivines and pyroxenes [14].

**Table 1.** Spectral parameters of luminescence for chondrites.

| Chondrite | Method | T, K | $E_{max}$, ± 0.05 eV | $\omega_E$, ± 0.05 eV | Reference |
|---|---|---|---|---|---|
| **Chelyabinsk** | TL | 393 | 2.81<br>2.43 | 0.68<br>0.40 | This work |
| | PL | RT | 2.80<br>2.45 | 0.70<br>0.37 | [6] |
| | CL | RT | 2.68 | 0.75 | [15] |
| **Tsarev** | TL | 368 | 2.81<br>2.47 | 0.70<br>0.42 | This work |
| | | 393 | 2.85<br>2.50 | 0.72<br>0.44 | |
| | | 418 | 2.87<br>2.50 | 0.73<br>0.42 | |
| | | 443 | 2.87<br>2.48 | 0.74<br>0.33 | |
| | PL | RT | 2.90<br>2.48 | 0.85<br>0.42 | [9] |

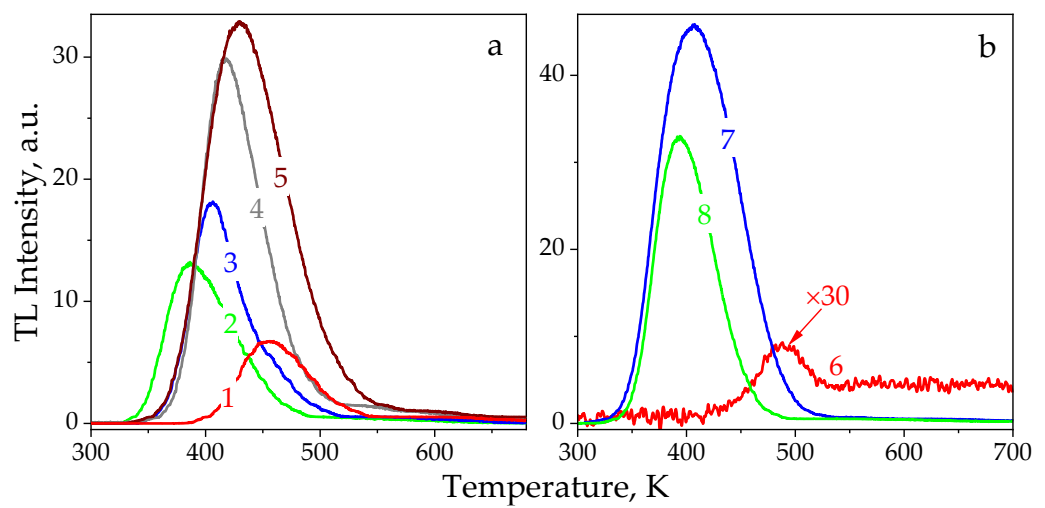

**Figure 4.** TL glow curves for meteorites irradiated with various doses:
(a) the Chelyabinsk chondrite: 1 – natural, 2 – 9.1 kGy, 3 – 18.2 kGy, 4 – 27.3 kGy, 5 – 36.4 kGy.
(b) the Tsarev chondrite: 6 – natural, 7 – 9.1 kGy, 8 – 9.1 kGy after 3 days of storage in dark.

Figure 4 shows natural and induced TL curves for the studied samples. As for the Chelyabinsk LL5 chondrite, a maximum of the natural TL is visible at 400 – 520 K. Apart from it, a high-temperature shoulder at 520 – 750 K was revealed. It is worth underscored that the Dhajala meteorite showcases similar parameters for the main TL peak [5]. The emission demonstrates two-ten-fold intensity changes; the maximum temperature of the TL peak shifts from $T_{max}$ = 427 to 487 K for different samples. It can be due to inhomogeneous mineral phases or mineral compositions, or differences in irradiation space doses [4]. The estimated values of shape parameters for the TL curves measured for the chondrites under study are presented in Table 2. It can be noted that the high temperature shift of the induced TL peaks with dose increasing is observed within the interval of $T_{max}$ = (390 – 430) ± 10 K, while the halfwidths change at $\omega_T$ = 61 – 83 K. The maximum temperatures of the natural TL peaks are noticeably higher for both meteorites.

The observed dependencies of the TL curves parameters on the irradiation dose cannot be described under the assumption made for independent charge capturing centers, particularly in the frame of the «one trap – one recombination center» model [16]. The high values calculated for a geometric factor of $\mu_g$ = 0.58 - 0.64 indicate the processes with the kinetics order of b > 2 and are consistent with the estimates performed earlier in [17]. These facts evince the possible presence of a quasi-continuous system of capturing levels, which are active and interact in the investigated temperature range. Such a situation is quite typical for silicates (pyroxene, olivine, and others), in which various structural defects form luminescent complexes to be responsible for the processes analyzed.

**Table 2.** TL parameters of chondrites after irradiation.

| Chondrite | Dose, kGy | $T_{max}$, K | $\omega_T$, K | $\mu_g$ |
|---|---|---|---|---|
| **Chelyabinsk** | 9.1 | 387 | 70 | 0.64 |
| | 18.2 | 402 | 66 | 0.62 |
| | 27.3 | 418 | 61 | 0.59 |
| | 36.4 | 429 | 83 | 0.58 |
| | natural | 456 | 72 | 0.56 |
| **Tsarev** | 9.1 | 408 | 86 | 0.53 |
| | 9.1 [1] | 394 | 62 | 0.58 |
| | natural | 490 | 60 | 0.50 |

[1] after 3 days of storage in dark

The natural TL glow curve for the Tsarev L5 chondrite was found to contain a low intensity peak with a maximum temperature $T_{max}$ = 490 K and halfwidth of $w_T$ = 60 K (see Figure 4b and Table 2). In addition, a storage time affects the induced TL maximum; it shifts to lower temperatures from 408 ± 5 to 394 ± 5 K. In this case, the half-width narrows from 86 ± 5 to 62 ± 5 K. The observed TL faded away 45 % after a 3-day holding period. For drawing a more reliable conclusion concerning the number of different capturing centers emptied within the RT – 500 K temperature range, it is necessary to use additional TL techniques such as dose or heating rate variation, step pre-heating etc. [2,18].

In the case of measuring the natural TL, the equilibrium signal with information about the accumulated dose is assumed to be read from a trap - the last emptying and the deepest one. Accordingly, natural glow curves obtained were analyzed in terms of the peak shape formalism for the general order kinetics [18]:

$$E_A = \left[ 0.976 + 7.3(\mu_g - 0.42) \right] \frac{kT_{max}^2}{\delta}, \qquad (1)$$

Here $\mu_g$ is the geometrical factor; $\delta$ is the high temperature half-width of the TL peak, k is the Boltzmann constant, J.K$^{-1}$ and $E_A$ is the activation energy in eV. The values of $E_A$ = 0.86 ± 0.10 and 1.08 ± 0.10 eV were calculated using the natural TL curves for the Chelyabinsk and Tsarev chondrites, respectively. The data obtained are in satisfactory agreement with $E_A$ = 0.9 – 1.6 eV taken from an analysis of the Dhajala meteorite thermoluminescence [5]. For the induced TL approximation, a superposition of several glow peaks which characterized the presence of a quasi-continuous system of traps should be used. We have no information enough to choose the number of the kinetic components.

## 4. Conclusions

The emission spectra, natural and induced TL glow curves in the 300 – 650 nm range were measured for the Chelyabinsk and Tsarev stony meteorites using a luminescent spectrometer with a developed high-temperature appliance. All the TL spectra were approximated by a superposition of two Gaussians with maximum energies near 2.8 and 2.5 eV. The 2.8 eV band dominates in the TL emission and has intensity 4 – 6 times higher than that of the 2.5 eV band. The conducted analysis of the obtained and independent data on spectral parameters of PL and CL allows one to conclude that the observed features of the luminescence can be caused by defective recombination centers in the structure of forsterite in the meteorite composition.

The study of induced TL has shown that a high-temperature shift of $\approx$ 40 K is observed for the TL peak maximum as dose increases within the 9.1 – 36.4 kGy range. We have revealed a quasi-continuous traps distribution being active at 350 – 650 K. Based on the general order kinetics and using natural TL data for the Chelyabinsk and Tsarev meteorites, we have also estimated the values of activation energies $E_A$ = 0.86 ± 0.10 and 1.08 ± 0.10 eV, respectively. This work has demonstrated that thermoluminescence processes in the Chelyabinsk LL5 and Tsarev L5 chondrites are characterized by similar spectral and kinetic peculiarities.

**Author Contributions:** Conceptualization, I.W. and A.V.; formal analysis, A.H.; investigation, A.V. and A.H.; resources, I.W. and O.H.; writing—original draft preparation, A.V. and A.H.; writing—review and editing, I.W., T.S. and O.H.; visualization, A.V. and I.W.; supervision, I.W. and T.S.; funding acquisition, I.W. and T.S. All authors have read and agreed to the published version of the manuscript.

**Funding:** This work was supported by Minobrnauki research project FEUZ-2020-0059.

**Acknowledgments:** Authors thank Dr. O. Ryabukhin and Dr. A. Ishchenko for help in irradiation of the samples. Fragment of the Tsarev chondrite has been provided for study by Dr. V. Grokhovsky.

**Conflicts of Interest:** The authors declare no conflict of interest. The funder had no role in the design of the study.